\documentclass[a4paper,12pt]{article}
\usepackage[english]{babel}
\usepackage[latin1]{inputenc}
\usepackage{amsthm,amsfonts,amssymb,amsmath}
\usepackage{graphicx,color}
\usepackage{subfigure}
\usepackage{textcomp}
\usepackage{multirow}
\usepackage{mathrsfs}
\usepackage[normalem]{ulem}
\usepackage{indentfirst}
\usepackage{geometry}
\usepackage{float}

\newtheorem{defi}{Definition}

\begin{document}
\begin{center}
\textsc{Anomalous relaxation in dielectrics with Hilfer fractional derivative}
\end{center}

\bigskip

\begin{center}
\begin{tabular}{cc}
{\sf A. R. G\'omez Plata} &{\sf Ester C. A. F. Rosa} \\
Department of Mathematics & Department of Applied Mathematics \\
Universidad Militar Nueva Granada & Imecc -- Unicamp \\
250247, Cajic\'a-Zipaquir\'a, Colombia & 13083-859, Campinas, SP, Brazil  \\
\end{tabular}

and

\begin{tabular}{cc}
{\sf R.G Rodriguez-Giraldo} &{\sf E. Capelas de Oliveira} \\
Department of Mathematics & Department of Applied Mathematics \\
Universidad Militar Nueva Granada & Imecc -- Unicamp \\
250247, Cajic\'a-Zipaquir\'a, Colombia & 13083-859, Campinas, SP, Brazil  \\
\end{tabular}
\end{center}

\begin{abstract}
We introduce a new relaxation function depending on an arbitrary parameter as solution of a kinetic equation
in the same way as the relaxation function introduced empirically by Debye, Cole-Cole, Davidson-Cole and Havriliak-Negami, anomalous
relaxation in dielectrics, which are recovered as particular cases. We propose a differential equation introducing a
fractional operator written in terms of the Hilfer fractional derivative of order $\xi$, with $0 < \xi \leq 1$ and type $\eta$, with $0 \leq \eta \leq 1$.
To discuss the solution of the fractional differential equation, the methodology of Laplace transform is required. As a by product we
mention particular cases where the solution is completely monotone. 
Finally, the empirical models are recovered as particular cases. 
\end{abstract}

\section{Introduction}
Fractional Calculus (FC) has greatly developed during the last years and has established itself as a generalization of classical differential
and integral calculus, as proposed by Newton and Leibniz, independentely. Seems to be clear that several phenomena, inherent to complex systems,
are expressed adequately by the theory of anomalous relaxation (anomalous diffusion, also) and the corresponding mathematical tools, incorporated in FC, for example
anomalous relaxation in dielectrics discussed in terms of the Mittag-Leffler functions \cite{ecomainardivaz,ecomainardivaz1}.
It is important to note that there are several different ways to introduce a fractional differential operator \cite{capelastenreiro}.
Also, in recent years many researches introduce new classes of fractional differential operators but, as we known, should not be considered as
fractional \cite{tarasov,tarasov1}. We mention two classes of them, as they appear in the specialized literature, named: the fractional local
operators and fractional derivative operator whose kernel is a non singular function. Here, after that mention, we refer to such operators as
local differential operator \cite{akkurt} and non singular kernel differential operator \cite{caputofabrizio,losada}. The first one can be shown
to be a multiple of the integer first order derivative and the second is not associated with the memory effect.

Further, FC is a research topic that plays a fundamental role and can be referred to as an important branch of so-called complexity science
which, among others, covers topics of various fronts, some of them being: applied mathematics, statistical mechanics, economics, etc.
For example, an important feature of the dynamics of supercooled liquids and amorphous polymers reflects the fact that the relaxation is not
of the exponential type, which emerges in several experiments: dielectric spectroscopy, measurements of the viscoelasticity modulus, dispersion
of quasielastic light, modulus of shear and shear compliance, among others. A non-exponential relaxation characterizes a particular deviation of
the Debye type relaxation function \cite{bohmer,hilfer2003,sabatier-agrawal-tenreiro,mainardi2010}. Let us focus on the relaxation properties of dielectric
materials which are described, in the frequency domain, through several models, all of
them recovering, as a particular case, the Debye model \cite{debye}. Here, just to mention, the models most studied in the literature, those that,
in the course of time, came with the intention of improving the classic Debye model, thus describing a particular phenomenon:
Kohlrausch-Williams-Watts \cite{kohlrausch,willians,cardona}, Cole-Cole \cite{colecole}, Davidson-Cole, \cite{coledavidson}, Havriliak-Negami
\cite{havriliak} and excess wing model \cite{casalini}.

Furthermore, in \cite{estereco}, a study was proposed discussing differential equations associated with the aforementioned models, whose solutions were given
in terms of Mittag-Leffler functions, in \cite{mainardigarrappa} discuss the complete monotonicity of the so-called Prabhakar function,
while in \cite{garrapamainardimaione} it was emphasized the character of the relaxation functions to be completely monotone,
as well as a characterization of the models in terms of fractional order differential operators. Also, in \cite{ester1} the fractional kinetic relaxation functions,
derived from fractional differential equations, with the fractional Riemann-Liouville differential operator, were discussed, aiming at its complete monotonicity.
On the other hand, we emphasize that during the last five decades, after the first congress dedicated exclusively to the FC, remarkable advances were made
in relation to investigations involving phenomena associated with anomalous relaxation and diffusion, as well as in the study of
fractional differential equations, only to stay in the themes of interest in the present research. For other issues and possible open issues, we mention the
discussions that took place at two round tables during the two International Conferences \emph{Fractional Differentiation and Applications}, \cite{table1,table2}.

It is important to mention a few recent papers on the subject of FC, involving anomalous relaxation and diffusion: a particular study on viscoelasticity by means of the
Prabhakar-like operator \cite{GiustiColombaro}; a time fractional convection-diffusion equation to model a particular gas transport \cite{Changetal};
a new property of the fractional Kuramoto model \cite{SeungJung}; an interesting discussion on the Havriliak-Negami relaxation model \cite{Gorskaetal};
a nice review of FC applications to the real world problems \cite{Sunetal} and a discussion on a general fractional relaxation equation \cite{EcoStefaniaVaz}.

As we have already mentioned, there are situations where Debye's theory does not apply, that is, the phenomenon of relaxation is not of the type
exponential, in particular, we mention amorphous polymers. The first work that investigated this fact was proposed in the 1940s \cite{colecole}
where an empirical formula was presented to describe the phenomenon. From this first empirical formula, the first attempt to correct the Debye model,
always adding a particular parameter, other models were proposed, among them the Davidson-Cole \cite{coledavidson} and Havriliak-Negami \cite{havriliak} models,
the latter a combination of the two previous ones. Note that in such proposals empirical equations were obtained, the so-called kinetic equations describing
relaxation processes in dielectrics, for the construction of a general microscopic theory is still an open problem, in particular, lacking a physical interpretation
involving the parameters introduced with the dynamic parameters of the medium. In this sense, a formulation of so-called kinetic equations with memory was proposed,
based on the Mori-Zwanzig formalism \cite{mori,zwanzig}.

It is important to note that, a simple question arises: how to justify the empirical laws? So, there are some different ways to answer this question.
Here, we mention some of them. First, Hilfer \cite{hilfer2000} put together Caputo and Riemann-Liouville fractional derivatives, introducing the
generalized fractional derivative of order $\xi$, $0 < \xi < 1$ and type $\eta$, $0 \leq \eta \leq 1$. Generalized fractional relaxation equations,
based on generalized Riemann-Liouville derivatives, were proposed by Hilfer \cite{hilfer2003} and more, with a simple short time regularization,
the equations were solved exactly.  An approach involving anomalous relaxation in dielectrics with fractional derivatives was proposed by
Novikov et al. \cite{novikov}. In this paper, the authors show that anomalous relaxation in dielectrics can be described in terms of the differential
equation with Riemann-Liouville fractional derivative. Also, the solutions of these equations are presented in terms of the Fox's $H$-function and
some particular cases in terms of the Mittag-Leffler function. Also, we mention Weron and collaborators \cite{weron,weron1} use the continuous time random walk method
but the most acceptable approach is due to the Mori-Zwanzig projection operators formalism because it is based on the general Hamiltonian formalism.
Recent, using the memory function formalism, as proposed by Mori-Zwanzig, Khamzin et al. \cite{khamzin} discuss the kinetic equations for relaxation functions
associated with the empirical function, for Davidson-Cole and Havriliak-Negami models. Cole-Cole and Debye models are recovered as particular
cases. A physical interpretation for Havriliak-Negami model is also given. In a more recent paper, Pandey \cite{pandey} discusses the anomalous relaxation
in dielectrics by means of the Lorenzo-Hartley's function and uses the Laplace transform to discuss the conditions that this function must satisfy to be completely monotone.

In this paper we will propose a new model for the anomalous relaxation in dielectrics, using only the so-called classical fractional derivative operator,
specifically the Hilfer fractional derivative operator of order $\xi$, with $0 < \xi \leq 1$ and type $\eta$, with $0 \leq \eta \leq 1$ and introducing a new real parameter. As
particular cases we will recover the classic models already mentioned.

The paper is organized as follows: in Section 2, we present some preliminaires, involving some remarks on the FC and the essential of the Mittag-Leffler
function, specifically the so-called Prabhakar-type function; in Section 3, we review the classical relaxation models, particularly, Havriliak-Negami and its
particular cases Davidson-Cole, Cole-Cole and Debye and, discussing kinetic equation modelling the empirical Havriliak-Negami model and obtaining it respective solution;
in Section 4, our main result, we introduce a general model, proposing a fractional differential equation with the Hilfer fractional derivative, presenting it
solution, in terms of the Prabhakar-type function, and specifically well-known models, are recovered. Concluding remarks close the paper.

\section{Preliminaires}
In this section, we introduce some basic notations and preliminary facts, presented as definitions, that will be used throughout the paper, specifically, Riemann-Liouville
fractional integral, Hilfer fractional derivative and Mittag-Leffler functions. There are several ways to introduce the concept of fractional integral.
We chose to introduce the concept of fractional integral using the Gel'fand-Shilov function in order to express the integral of order $n$,
with $n\in\mathbb{R}$, as a Laplace convolution product. The factorial is replaced by a gamma function in order to obtain what is known as a fractional integral
in the Riemann-Liouville sense or simply Riemann-Liouville integral. It is from this fractional integral that the concept of Hilfer fractional derivative
will be introduced.

We conclude the section presenting a particular product of a power function and the Mittag-Leffler function with three parameters \cite{gorenflo},
sometimes known as Prabhakar-type function, its particular cases and the respective pair of Laplace transforms, direct and inverse \cite{rubeco}.

\begin{defi}
\sc{Gel'fand-Shilov function}
\end{defi}
Let $n \in \mathbb{N}$ and $\nu \in \mathbb{R}$. The Gel'fand-Shilov function, denoted by $\phi(t)$, is defined by the following expressions
$$
\phi_n(t):=\left\{ \begin{array}{ccl}
                    \displaystyle \frac{t^{n-1}}{(n-1)!} & {\rm if} & t \geq 0\\
                    &&\\
                    0 & {\rm if} & t < 0
                   \end{array} \right. \qquad {\rm and} \qquad
                   \phi_{\nu}(t):=\left\{ \begin{array}{ccl}
                    \displaystyle \frac{t^{\nu-1}}{\Gamma(\nu)} & {\rm if} & t \geq 0\\
                    &&\\
                    0 & {\rm if} & t < 0
                   \end{array} \right.
$$
respectively.

\begin{defi}
\sc{Integral of order $\nu$}
\end{defi}
Let $f(t)$ be a locally integrable function. We define the integral of a locally integrable function, $f(t)$, of order $\nu$, $\nu \in \mathbb{R}$, denoted by
$J^{\nu} f(t)$, by means of
$$
J^{\nu} f(t) := \phi_{\nu}(t) * f(t) = \frac{1}{\Gamma(\nu)} \int_0^t (t- \tau)^{\nu -1} f(t) \, {\mbox{d}}\tau
$$
with $*$ denotes the Laplace convolution product.

\begin{defi}
\sc{Riemann-Liouville fractional integral}
\end{defi}\label{LI}
Let $\Omega = [a,b]$ with $ - \infty < a < b < \infty$ be a finite interval on the real axis. Let $f(t)$ be a locally integrable function.
The Riemann-Liouville fractional integrals, denoted by, $(J_{a^{+}}^{\nu} f)(x) \equiv {}_aJ_x^{\nu} f(x)$ and $(J_{b^{-}}^{\nu} f)(x) \equiv {}_xJ_b^{\nu} f(x)$
both of order $\nu \in \mathbb{C}$, with Re$(\nu) > 0$ and $\nu \neq \mathbb{N}$, are defined by
$$
(J_{a^{+}}^{\nu} f)(x) := \frac{1}{\Gamma(\nu)} \int_a^x (x-t)^{\nu -1} f(t) \, {\mbox{d}}t, \qquad x > a, \qquad {\mbox{Re}}(\nu) > 0
$$
and
$$
(J_{b^{-}}^{\nu} f)(x) := \frac{1}{\Gamma(\nu)} \int_x^b (t - x)^{\nu -1} f(t) \, {\mbox{d}}t, \qquad x < b, \qquad {\mbox{Re}}(\nu) > 0
$$
respectively.

These fractional integrals are called Riemann-Liouville fractional integrals on the left and right, respectively.
Note that in the case where $\nu = n\in\mathbb{N}$ these definitions coincide with integer order integrals.

\begin{defi}
\sc{Hilfer fractional derivative}\label{H}
\end{defi}
Let $\Omega = [a,b]$ with $ - \infty < a < b < \infty$ be a finite interval on the real axis. Let $f(t)$ be a locally integrable function.
The Hilfer (right-/left-sided) fractional derivatives, denoted by, $(\mathscr{D}_{a^{+}}^{\xi,\eta} f)(t)$ and $(\mathscr{D}_{b^{-}}^{\xi,\eta} f)(t)$,
both of order $\xi \in \mathbb{C}$, with Re$(\xi) > 0$, and type $\eta$, with $0 \leq \eta \leq 1$, are defined by
$$
\begin{array}{rcl}
(\mathscr{D}_{a^{+}}^{\xi,\eta} f)(t) &:=& \displaystyle \left[ J_{a^{+}}^{\eta(1-\xi)} \frac{{\mbox{d}}}{{\mbox{d}}t} \left( J_{a^{+}}^{(1-\eta)(1-\xi)} f \right) \right](t)\\
\end{array}
$$
and
$$
\begin{array}{rcl}
(\mathscr{D}_{b^{-}}^{\xi,\eta} f)(t) &:=& - \displaystyle \left[ J_{b^{-}}^{\eta(1-\xi)} \frac{{\mbox{d}}}{{\mbox{d}}t} \left( J_{b^{-}}^{(1-\eta)(1-\xi)} f \right) \right](t)\\
\end{array}
$$
respectively, with $J_{{c}^{\pm}}^{\nu}$ are the Riemann-Liouville integrals, given by Eq.(\ref{LI}).



Two particular cases of the type are well-known. In the case $a > -\infty$ and type $\eta = 0$, we recover the classical Riemann-Liouville fractional
derivative, an integer-order derivative of a fractional integral, while for $\eta = 1$, we recover the classical Caputo fractional derivative,
a fractional integral of an integer-order derivative \cite{hilfer2000}.

\begin{defi}
\sc{Laplace transform and Hilfer fractional derivative}
\end{defi}\label{HLT}

Let $s$ be the parameter of the Laplace transform with Re$(s)>0$. The Laplace transform, denote by $\mathscr{L}$, of the Hilfer fractional derivative is given by
\begin{equation}
\mathscr{L}[\mathscr{D}_{a^{+}}^{\xi,\eta} f(t)](s) = s^{\xi} F(s) - s^{\eta(\xi-1)} \left[ \mathscr{D}_{a^{+}}^{(1-\eta)(\xi-1),0}f \right] (0^{+})
\end{equation}
with the notation $F(s):=\mathscr{L}[f(t)]$, where the initial condition is the Riemann-Liouville derivative evaluate at $t \to 0^{+}$. For the type
$\eta = 0$, the fractional derivatives involves fractional initial values while for $\eta =1$ the initial values are nonfractional.

\begin{defi}
\sc{Mittag-Leffler function with three parameters}
\end{defi}
Let $\alpha,\beta,\rho \in \mathbb{C}$, with Re$(\alpha) > 0$, $z \in \mathbb{C}$ and $(\rho)_k$ the Pochhammer symbol. The Mittag-Leffler function with three parameters,
denoted by, $E_{\alpha,\beta}^{\rho}(\cdot)$, is defined by \cite{gorenflo}
\begin{equation}\label{00}
E_{\alpha,\beta}^{\rho}(z) = \sum_{k=0}^{\infty} \frac{(\rho)_k}{\Gamma(\alpha k + \beta)} \frac{z^k}{k!} \cdot
\end{equation}
Note that, $E_{\alpha,\beta}^{1}(z) = E_{\alpha,\beta}(z)$ is the Mittag-Leffler function with two parameter and $E_{\alpha,1}^{1}(z) = E_{\alpha}(z)$ is
the classical Mittag-Leffler function introduced by himself \cite{mittag}.

We also introduce a particular function involving the Mittag-Leffler function, sometimes known called Prabhakar-type function, specifically, a product of a
power function and the Mittag-Leffler function with three parameters, by means of
\begin{equation}\label{E}
\mathscr{E}_{\alpha,\beta}^{\rho} (t) := t^{\beta-1} E_{\alpha,\beta}^{\rho}(t),
\end{equation}
with the restrictions as in the Mittag-Leffler function with three parameters, Eq.(\ref{00}).

\begin{defi}
\sc{Direct and inverse Laplace transform of $\mathscr{E}_{\alpha,\beta}^{\rho} (t)$}
\end{defi}
Let $\alpha,\beta,\rho \in \mathbb{C}$, with Re$(\alpha) > 0$, $t > 0$ and $a \in \mathbb{R}^{*}$. The Laplace transform, denoted by $\mathscr{L}$, of the
$\mathscr{E}_{a,b}^c (t)$ is given by
$$
\mathscr{L} [\mathscr{E}_{\alpha,\beta}^{\rho}(t)] \equiv \mathscr{L}\left[t^{\beta -1} E_{\alpha,\beta}^{\rho} (a t^{\alpha}) \right] = \frac{s^{\alpha \rho - \beta}}{(s^{\alpha}-a)^{\rho}}
$$
where $s \in \mathbb{C}$ is the parameter of the Laplace transform. Also, denoting the inverse Laplace transform by $\mathscr{L}^{-1}$, we can write
$$
\mathscr{L}^{-1}\left[ \frac{s^{\alpha \rho - \beta}}{(s^{\alpha}-a)^{\rho}} \right] = t^{\beta -1} E_{\alpha,\beta}^{\rho} (a t^{\alpha}) \equiv \mathscr{E}_{\alpha,\beta}^{\rho}(t).
$$
As we have already said, taking a particular value of the parameters, we obtain the pair (direct and inverse) of Laplace transform associated
with the Mittag-Leffler function with two and one parameter.

\section{Classical relaxation models}
The classical relaxation models, also known as empirical models, specifically associated with the dielectrics, appear after the classical
Debye model, proposed in 1929, as a deviation of the Debye model. In this section we will present a brief summary of the paper \cite{estereco},
but presenting only the Havriliak-Negami model because it is the most general of the three mentioned above. Note that, as particular cases of
the Havriliak-Negami model we can recover the other two models, Davidson-Cole and Cole-Cole and, also, the classical Debye model.
Here, we refer these four models as classical relaxation models, or classical empirical models.

The most important phenomenon associated with a dielectric material is its polarization. The time lapse necessary for the material
to respond to the applied electric field is called relaxation time. Thus, when submitted to an electric excitation, the dielectric will
respond to this action in an attempt to reestablish its equilibrium during and after the electric stimulus. The polarization of each dielectric material
depends on the nature of its molecular and atomic chemical bonds, and there is presently no univeral model which can explain the polarization
phenomenon in all materials \cite{bottcher}.

Debye's response function was the first theoretical model for the dielectric behaviour of some substances \cite{debye}. However, due to its
limitations, this model is incapable of describing in details the dielectric response of a large number of solids and liquids. After this Debye's theory,
several other response functions were proposed to serve as models for describing the dielectric relaxation. Among then, we here focus our attention on the
model proposed by Cole-Cole \cite{colecole} and the one by Davidson-Cole \cite{coledavidson}, both of which emerged in attempts to adjust the
response function to the experimental behaviour of some dielectric materials. There is also the model of Havriliak-Negami \cite{havriliak}
which can be considered as a generalization of the latter models.

These last three models, called anomalous models, are the most relevant
in the literature; however, there are other models which approach the phenomena from a different perspective, as the model by Hilfer \cite{hilfer2003}, Weron and
collaborators \cite{weron,weron1} and Hanyga-Seredynska \cite{hanyga}.

\subsection{Havriliak-Negami relaxation model}
In their works, Havriliak-Negami \cite{havriliak,havriliak1} studied the complex dielectric behaviour of twenty-one polymers and noticed that they had approximately
the same form. They then arrived at the following empirical expression for the complex susceptibility, representing the relaxation process,
\begin{equation}\label{HN1}
\tilde{\varepsilon}_{{\sf{HN}}}(s) = \frac{1}{(1+(s\sigma)^{\alpha})^{\beta}}
\end{equation}
where $\tilde{\varepsilon}_{{\sf{HN}}}$ is the complex susceptibility and $\sigma$ the constant associated with the dipole's characteristic relaxation time.
The real constants $\alpha$ and $\beta$ satisfying the inequalities $0 < \alpha \leq 1$ and $0 < \beta \leq 1$ and $s$ a parameter associated with
the Laplace transform in variable $s=i\omega$. It is important to observe that for $\beta = 1$ we have the Cole-Cole model, while $\alpha = 1$
takes to Davidson-Cole model and, finally, with $\alpha = \beta = 1$ we recover the first model as proposed by Debye \cite{estereco}.

\subsection{Kinetic equation}
The properties of dielectric materials are usually described by two real constants which are called dielectric constants. They can be combined
in a complex dielectric constant, known as complex dielectric permitivity, which is given by the following superposition relation \cite{frohlich}
\begin{equation}\label{HN2}
\tilde{\varepsilon}(i\omega) = \mathscr{L}\left[- \frac{{\mbox{d}}\varphi(t)}{{\mbox{d}}t} \right] (i\omega) = 1 - i \omega \cdot \tilde{\varphi}(i\omega)
\end{equation}
where $\mathscr{L}[f(t)](i\omega)=\tilde{f}(i\omega)$ is the Laplace transform of $f(t)$ in variable $i\omega$. We have that $\varphi(t)$ is the
normalized polarization decay function when a macroscopic electric field is removed from its medium. Function $\varphi(t)$ contains only the
contributions from the relaxation process and we have chosen $\varphi(0)=1$ \cite{manning}.

It is important to note that, in the case of linear approximation response, the polarization changes caused by thermal
motion are the same as for the macroscopic function dipole relaxation induced by the electric field. Therefore, the laws governing the
dipole correlation function, denoted by $\phi(t)$, are directly related to the kinetic properties and macroscopic structures of the dielectric system,
represented by function $\varphi(t)$. Thus, it is possible to equate the relaxation function $\varphi(t)$ to the macroscopic dipole correlation
function $\phi(t)$ as follows
$$
\varphi(t) \simeq \phi(t) = \frac{\langle M(t) M(0)\rangle}{\langle M(0) M(0) \rangle}
$$
where $M(t)$ is the fluctuating macroscopic dipole moment \cite{willians1}.

As we have already said, the memory function formalism, as proposed by Mori-Zwanzig \cite{mori,zwanzig}, is the most acceptable approach
to discuss kinetic equations for relaxation functions associated with the empirical function.  Also, we have mentioned Khamzin et al. \cite{khamzin}
discussed the kinetic equations for relaxation functions associated with the classical empirical functions.

Here, we will review this formalism in order to obtain the equation regarding the Havriliak-Negami model as well as the respective solution.
We note that, Debye, Cole-Cole and Davidson-Cole models can also be obtained as particular cases. First, we introduce the function $\phi(t)$,
temporal correlation function, whose dipole correlation function as defined above is a specific case. Thus, this function is a solution of an
integro-differential equation \cite{boon}
\begin{equation}\label{HN3}
\frac{{\mbox{d}}\phi(t)}{{\mbox{d}}t} = - \int_0^t K(t-\xi) \, \phi(\xi) \, {\mbox{d}}\xi .
\end{equation}
which takes into account the effects of memory. Therefore, introducing the concept of an integral memory function given
by $M(t) = \displaystyle \int_0^t K(\xi) \, {\mbox{d}}\xi$ and using the fact that, in linear approximation response, the relaxation function
$\varphi(t)$ also satisfies Eq.(\ref{HN3}), we obtain the following relation
\begin{equation}\label{HN4}
\frac{{\mbox{d}}\varphi(t)}{{\mbox{d}}t} = - \frac{{\mbox{d}}}{{\mbox{d}}t} \int_0^t M(t - \xi) \, \varphi(\xi) \, {\mbox{d}}\xi =
- \frac{{\mbox{d}}}{{\mbox{d}}t} M(t) * \varphi(t)
\end{equation}
where $*$ denotes a convolution product. Using Eq.(\ref{HN1}) and Eq.(\ref{HN4}) and evaluating the Laplace transform on both members of Eq.(\ref{HN4})
we obtain the integral memory function, $M(t)$, given in terms of the inverse Laplace transform of $M(t)$, denoted by $\tilde{M}(i\omega)$,
\begin{equation}\label{HN5}
\tilde{\varepsilon}(i\omega) = \frac{1}{1+\tilde{M}^{-1}(i\omega)} \cdot
\end{equation}

Comparing the relation expressed by Eq.(\ref{HN5}) with the classical empirical law, associated with the Havriliak-Negami model, Eq.(\ref{HN1}),
with $s=i\omega$ we obtain, after taking the inverse Laplace transform, the corresponding memory function in the time variable
\begin{equation}\label{HN6}
M_{{\sf{HN}}}(t) = \frac{1}{t} \sum_{k=0}^{\infty} \left(\frac{t}{\sigma} \right)^{\alpha \beta (k+1)}
E_{\alpha,\alpha \beta(k+1)}^{\beta(k+1)} \left[- \left( \frac{t}{\sigma} \right)^{\alpha}  \right]
\end{equation}
with $0 < \alpha \leq 1$, $0 < \beta \leq 1$ and where $E_{a,b}^c (\cdot)$ is the Mittag-Leffler function with three parameters as given in Eq.(\ref{00}).
We also remember that, taking particular values of the parameters $\alpha$ and $\beta$ we obtain the corresponding memory function in the
time variable for the Debye, Cole-Cole and Davidson-Cole models \cite{estereco}.

Substituting the memory function in the time variable given by Eq.({\ref{HN6}) into Eq.(\ref{HN4}) we obtain the kinetic equation associated with
the Havriliak-Negami model
\begin{equation}\label{HN7}
\frac{{\mbox{d}}}{{\mbox{d}}t} \left\{\varphi(t) + \sum_{k=0}^{\infty} \int_0^t \frac{(t-\xi)^{\alpha \beta(k+1)-1}}{\sigma^{\alpha \beta(k+1)}}
E_{\alpha,\alpha\beta(k+1)}^{\beta(k+1)} \left[-\left( \frac{t-\xi}{\sigma}  \right)^{\alpha}  \right] \, \varphi(\xi) \, {\mbox{d}}\xi  \right\} = 0
\end{equation}
whose solution, using the initial condition $\varphi(0)=1$, is given by
\begin{equation}\label{HN8}
\varphi_{{\sf{HN}}}(t) = 1 - \left( \frac{t}{\sigma} \right)^{\alpha \beta} E_{\alpha,\alpha\beta+1}^{\beta}
\left[-\left(\frac{t}{\sigma}\right)^{\alpha} \right]
\end{equation}
which is the relaxation function associated with the Havriliak-Negami model. Taking particular values of the parameters $\alpha$ and $\beta$ we obtain the corresponding
relaxation function for the Debye, Cole-Cole and Davidson-Cole models \cite{estereco}.

To conclude this section we mention that, in \cite{estereco} the authors introduce the Riemann-Liouville fractional derivative to discuss the classical
dielectric models. They introduce a parameter $\gamma$, with $(0,1]$, the order of the derivative and using the same procedure as above to obtain a
fractional differential equation in which the derivative is considered as in {\bf Definition \ref{H}.}, with the type $\eta = 0$, whose solution, the fractional relaxation function
associated with the fractional Havriliak-Negami model, $\varphi_{{\sf{FHN}}}(t)$, is given by
\begin{equation}\label{HN9}
\varphi_{{\sf{FHN}}}(t) = \frac{t^{\gamma-1}}{\Gamma(\gamma)} - \sigma^{-\alpha \beta} t^{\alpha \beta + \gamma -1} E_{\alpha,\alpha\beta+\gamma}^{\beta}
\left[-\left(\frac{t}{\sigma}\right)^{\alpha} \right],
\end{equation}
with the parameters $0 < \alpha \leq 1$, $0 < \beta \leq 1$, $0 < \gamma \leq 1$, $\sigma$ is a constant associated with the dipole's characteristic relaxation time and
$E_{a,b}^c (\cdot)$ is the Mittag-Leffler function with three parameters.

Note that, taking $\gamma=1$ we recover Eq.(\ref{HN8}) and, as above, the empirical Debye, Cole-Cole and Davidson-Cole models are recovered for particular values
of the parameters $\alpha$ and $\beta$. Finally, the paper \cite{estereco} presents several figures in which the asymptotic behavior of the function is studied.
The complex dielectric permittivity with the Riemann-Liouville fractional derivative is also discussed.

\section{General fractional relaxation model}
In this section, our main result, we propose a general fractional relaxation model with the Hilfer fractional derivative,
of order $0 < \xi < 1$ and type $0 \leq \eta \leq 1$, as introduced in {\bf Definition \ref{H}.} For this model, using the same procedure as above, we obtain a fractional
differential equation whose solution is the so-called general fractional relaxation, denoted by $\varphi_{\sf{GH}}(t)$. To this end we introduce a new parameter,
besides those that are present in the fractional Havriliak-Negami model. All previous particular cases are recovered.

Let $\mu$, $\alpha$ and $\beta$ be real parameters, with conditions to be imposed so that we can discuss the complete monotonicity of the
solution of the fractional differential equation. Thus, we propose the following expression for the complex susceptibility, $\tilde{\varepsilon}_{\sf{GH}}(s)$,
$$
\tilde{\varepsilon}_{\sf{GH}}(s) = \frac{(s\sigma)^{\mu-1}}{(1+(s\sigma)^\alpha)^{\beta}}
$$
being $s$ the parameter associated with the Laplace transform and $\sigma$ the constant associated with the dipole's characteristic relaxation time.
With this function, we will obtain the memory function in the time domain in order to write the respective fractional differential equation using the
Hilfer fractional derivative. Solve the fractional differential equation and mention the monotonicity of the solution. Recover as particular
cases known results associated with the fractional Havriliak-Negami, Davidson-Cole, Cole-Cole and Debye models and the classical empirical Havriliak-Negami,
Davidson-Cole, Cole-Cole and Debye models \cite{ecomainardivaz,ecomainardivaz1,ester1}. We will conjecture about a possible relation between the parameters of
the fractional derivative, order, $\xi$, and type, $\eta$, and the parameter $\mu$.

\subsection{Fractional kinetic equation}
First, proceeeding as before, Section 3.2, we obtain the corresponding memory function, i.e., using the complex susceptibility and Eq.(\ref{HN5}) we have
$$
\tilde{M}_{\sf{GH}}(s) = \frac{(\sigma s)^{\mu -1}}{(1+(\sigma s)^{\alpha})^{\beta} - (\sigma s)^{\mu -1}}
$$
whose inverse Laplace transforms provides
\begin{equation}\label{memoryf}
M_{\sf{GH}}(t) = \frac{1}{\sigma} \left(\frac{t}{\sigma} \right)^{\alpha \beta - \mu} \sum_{k=0}^{\infty} \left(\frac{t}{\sigma}\right)^{(\alpha\beta - \mu + 1)k}
E_{\alpha,(\alpha \beta - \mu +1)(k+1)}^{\beta(k+1)} \left[- \left(\frac{t}{\sigma}\right)^{\alpha} \right]
\end{equation}
with the parameters $0 < \alpha \leq 1$, $0 < \beta \leq 1$ and $E_{a,b}^c (\cdot)$ is the Mittag-Leffler function with three parameters as defined in Eq.(\ref{00}).

Second, similarly as in Eq.(\ref{HN4}), we will solve the following fractional differential equation
\begin{equation}\label{Hilfer}
\mathscr{D}_{a^{\pm}}^{\xi,\eta} \varphi_{\sf{GH}}(t) = - \mathscr{D}_{a^{\pm}}^{\xi,\eta} \left\{ M_{\sf{GH}}(t) * \varphi_{\sf{GH}}(t) \right\}
\end{equation}
where $\mathscr{D}_{a^{\pm}}^{\xi,\eta} (\cdot)$ is the Hilfer fractional derivative with order $\xi$, with $0 < \xi < 1$ and type $\eta$, with $0 \leq \eta \leq 1$.
Also, $*$ denotes the Laplace convolution product and $M_{\sf{GH}}(t)$, the memory function, is given by Eq.(\ref{memoryf}).

To solve Eq.(\ref{Hilfer}) we use the Laplace transform methodology, and for this end we will adopt as initial condition
$[\mathscr{D}_{a^{+}}^{(1-\eta)(\xi-1),0} \varphi(t)](0^{+})=1$ which is the Riemann-Liouville fractional derivative evaluated in $t \to 0^{+}$. Thus, taking the
Laplace transform of Eq.(\ref{Hilfer}) and using the initial condition, we have
$$
s^{\xi} \tilde{\varphi}(s) - s^{\eta(\xi-1)} = - \mathscr{L} \left[\mathscr{D}_{a}^{\xi,\eta} \left\{ M_{\sf{GH}}(t) * \varphi(t) \right\} \right]
$$
where $\tilde{\varphi}(s) \equiv \tilde{\varphi}_{\sf GH}(s)$ is the Laplace transform of $\varphi(t) \equiv {\varphi}_{\sf GH}(t)$ with $s$ the corresponding parameter. This algebraic equation can be written as follows
$$
s^{\xi} \tilde{\varphi}(s) - s^{\eta(\xi-1)} = - s^{\xi} \mathscr{L}[M_{\sf{GH}}(t)] \tilde{\varphi}(s)]
$$
whose solution is given by
\begin{equation}\label{HilferS}
\tilde{\varphi}(s) = \frac{s^{\eta(\xi-1)-\xi}}{1+ \mathscr{L}[M_{\sf{GH}}(t)]} \cdot
\end{equation}
Finally, substituting the expression for the Laplace transform of $M_{\sf{GH}}(t)$ in Eq.(\ref{HilferS}) and simplifying we have
$$
\tilde{\varphi}(s) = s^{\eta(\xi-1)-\xi} - \frac{s^{\mu -1 + \eta(\xi-1)-\xi}}{\sigma^{\alpha \beta - \mu +1} (s^{\alpha} + \sigma^{-\alpha})^{\beta}}
$$
whose corresponding inverse Laplace transform provides
\begin{equation}\label{Hilfer3}
\varphi(t) = \frac{t^{\gamma+\mu-\alpha \beta-1}}{\Gamma(\gamma+\mu - \alpha \beta)} - \frac{1}{\sigma^{\alpha \beta - \mu +1}} \mathscr{E}_{\alpha,\gamma+1}^{\beta}
\left[ - \left(\frac{t}{\sigma} \right)^{\alpha} \right]
\end{equation}
where we have introduced the parameter, which connect all four parameters
\begin{equation}\label{Parameter}
\gamma : = \alpha \beta + \xi -\eta(\xi-1) - \mu
\end{equation}
and $\mathscr{E}_{a,b}^c (\cdot)$ as defined in Eq.(\ref{E}).

Then, Eq.(\ref{Hilfer3}) with $\gamma$ given by Eq.(\ref{Parameter}) is the solution of Eq.(\ref{Hilfer}), i.e., the relaxation function for our general model.

\subsection{Particular cases}
In this section we will recover some known particular cases. First, relatively to the type of the Hilfer fractional derivative we have: (a) $\eta =1$,
the so-called Caputo type, we get $\gamma = \alpha - \mu + 1$, independently of the order; (b) $\eta =0$, the so-called Riemann-Liouville type, we get $\gamma = \alpha  - \mu + \xi$
which depends on the order. On the other hand, in the case $\mu=1$ we get $\gamma = \alpha + (1-\eta)(\xi-1)$, the relaxation function involves the Caputo type
($\eta=1$) and Riemann-Liouville type ($\eta=0$), both with the same conclusion as before, involving the dependency on the order.

In the case ($\eta=0$) with $\mu=1$ we recover the relaxation function associated with the fractional Havriliak-Negami model \cite{estereco}, as given in Eq.(\ref{HN9})
and also, $\xi=1$ we recover the classical Havriliak-Negami model \cite{havriliak,ester1}, as in Eq.(\ref{HN8}), and more, in both last two cases the corresponding fractional
and nonfractional models (classical anomalous models) particular cases are recovered: (i) $\alpha =1$ and $\beta \neq 1$, Davidson-Cole model; (ii) $\beta=1$ and $\alpha \neq 1$, Cole-Cole
model and, (iii) $\alpha=1=\beta$ the classical Debye model.

Below we will resume these particular cases, writting only the relaxation function, solution of the corresponding fractional differential equation, Eq.(\ref{Hilfer}).

\subsubsection{Riemann-Liouville type $\eta=0$}
Substituting $\eta = 0$ in Eq.(\ref{Hilfer3}), we have
$$
\varphi(t) = \frac{t^{\xi -1}}{\Gamma(\xi)} - \frac{1}{\sigma^{\alpha \beta - \mu +1}} \mathscr{E}_{\alpha,\alpha \beta + \xi - \mu + 1}^{\beta} \left[-\left(
\frac{t}{\sigma} \right)^{\alpha} \right]
$$
with $0 < \xi \leq 1$, $0 < \alpha \leq 1$, $0 < \beta \leq 1$ and $\mu$ a free parameter.

\subsubsection{Riemann-Liouville type $\eta=0$ and $\mu=1$}
Substituting $\eta = 0$ and $\mu=1$ in Eq.(\ref{Hilfer3}), we have
$$
\varphi_{\sf FHN}(t) = \frac{t^{\xi -1}}{\Gamma(\xi)} - \frac{1}{\sigma^{\alpha \beta}} \mathscr{E}_{\alpha,\alpha \beta + \xi}^{\beta} \left[-\left(
\frac{t}{\sigma} \right)^{\alpha} \right]
$$
with $0 < \xi \leq 1$, $0 < \alpha \leq 1$ and $0 < \beta \leq 1$. $\varphi_{\sf FHN}(t)$ is the relaxation function associated with the fractional Havriliak-Negami model
\cite{estereco} as given in Eq.(\ref{HN9}).

\subsubsection{Riemann-Liouville type $\eta=0$ and $\mu=1=\xi$}
Substituting $\eta = 0$ and $\mu=1=\xi$ in Eq.(\ref{Hilfer3}), we have
$$
\varphi_{\sf HN}(t) = 1 - \frac{1}{\sigma^{\alpha \beta}} \mathscr{E}_{\alpha,\alpha \beta + 1}^{\beta} \left[-\left(
\frac{t}{\sigma} \right)^{\alpha} \right]
$$
with $0 < \alpha \leq 1$ and $0 < \beta \leq 1$. $\varphi_{\sf HN}(t)$ is the relaxation function associated with the Havriliak-Negami model
\cite{havriliak,ester1} as given in Eq.(\ref{HN8}).

As we have already said, in both last two cases the corresponding fractional (Riemann-Liouville type $\eta=0$ and $\mu=1$)
and classical anomalous models (Riemann-Liouville type $\eta=0$ and $\mu=1=\xi$) models, particular cases are recovered:

\begin{center}
\begin{tabular}{rcl}
(i) & $\alpha =1$ and $\beta \neq 1$ & Davidson-Cole model, \\
(ii)& $\beta=1$ and $\alpha \neq 1$ & Cole-Cole model, \\
(iii)& $\alpha=1$ and $\beta=1$ & classical Debye model.
\end{tabular}
\end{center}

\subsubsection{Caputo type $\eta=1$ and $\mu=1$}
It is important to note that, in this case we recover directly the Havriliak-Negami model and its particular cases, Davidson-Cole,
Cole-Cole and Debye models \cite{havriliak,ester1}.

\section{Concluding remarks}
In this paper we proposed a general model to discuss relaxation associated with dielectrics. We used the Hilfer fractional derivative
and introduced a complex susceptibility depending on three parameters. By means of the Mori-Zwanzig formalism and the Laplace transform
methodology, we obtained the corresponding memory function in the time variable, given in terms of a Mittag-Leffler function with three
parameters, and obtained the respective fractional kinetic equation, as well as its solution, the so-called fractional relaxation function.

From this general model, we retrieve the formulations given in terms of the Riemann-Liouville and Caputo fractional derivatives as particular
cases of the parameters (order and type of Hilfer fractional derivative). For the Riemann-Liouville formulation we retrieved the fractional
relaxation function associated with the fractional Havriliak-Negami model as well as the fractional relaxation function associated with the
fractional models Davidson-Cole, Cole-Cole and Debye. Finally, the relaxation function associated with the classic Havriliak-Negami model
(Davidson-Cole, Cole-Cole and Debye, also) is retrieved when the order of the derivative is unitary.

We highlight the fact that the result involving the relaxation function associated with the classic Havriliak-Negami model (Davidson-Cole,
Cole-Cole and Debye, also) is also recovered through the Caputo formulation, a fact that we do not find in the literature.

We conclude by mentioning the importance of the complete monotonicity of the relaxation function, since it is a fundamental requirement
to be able to represent a physical quantity involving the so-called memory effect. Studies in this direction are in progress, in particular,
we conjecture that there must be a relation between the parameters so that we have a completely monotonous fractional relaxation function,
just as in the case where the derivative is unity \cite{RonaldAdrianEco}.

\subsection*{Acknowledgment}
ARGM is grateful, for the financial aid, of the Research Office-UMNG in the project IMP-CIAS 2651. This paper was completed when ECO was
visiting professor in the Department of Mathematics, Universidad Militar Nueva Granada (Campinas) as a recipient of a Fellowship of the IMP-CIAS 2651.

\end{document}